# The HTC-Claw: Automating Discovery through High-Throughput Computational Campaigns

*Lianduan Zeng, Xiao Zhou\*, Xueru Zheng, Ning Gao, Lei Liu, Yunxuan Cao, Hongjian Chen\*, Zhongyang Wang\*, Tongxiang Fan\**

L. Zeng, X. Zhou, L. Huang, L. Yang, L. Wang, Z. Wang, T. Fan

State Key Laboratory of Metal Matrix Composites, School of Materials Science and Engineering, Shanghai Jiao Tong University, Shanghai 200240, China.

E-mail: zhouxiao113@sjtu.edu.cn (Xiao Zhou); zy_wang@sjtu.edu.cn (Zhongyang Wang); txfan@sjtu.edu.cn (Tongxiang Fan)

## Abstract

With the advancement of the Materials Genome Initiative, high-throughput computation has become central to accelerating materials discovery. However, conventional first-principles workflows are cumbersome and error-prone. Existing high-throughput tools, while efficient at batch job submission, lack intelligence: they cannot automatically plan tasks based on scientific objectives or dynamically adapt workflows according to intermediate results. To address these limitations, this paper proposes and implements HTC-Claw, an intelligent high-throughput computational platform built upon the OpenClaw framework. The key innovations of HTC-Claw are: 1) An agent-based framework for automatic decomposition of high-level research goals into parallelizable task sets; 2) A closed-loop execution engine that integrates real-time analysis and reporting; 3) Adaptive decision-making and workflow iteration capabilities based on intermediate results; and 4) A decoupled, modular architecture that separates the scheduling system from functional modules, enhancing extensibility and robustness. Case studies demonstrate that HTC-Claw enables an intelligent, end-to-end workflow from user intent to final reporting in materials exploration

## 1. Introduction

With the rapid advancement of the Materials Genome Initiative, high-throughput computation has emerged as a central paradigm for accelerating materials discovery[1][2]. First-principles calculations, as an ab initio approach grounded in quantum mechanics, enable accurate prediction of key material properties, including electronic structure, mechanical behavior, and optical characteristics, thereby providing a solid theoretical foundation for rational materials design[3], However, conventional first-principles computational workflows involve multiple stages—such as structural optimization, parameter configuration, job submission, and result analysis—which are typically performed manually. This process is not only time-consuming and labor-intensive but also prone to human-induced errors.

In practical materials research, it is often necessary to perform systematic property screening across many candidate materials. For instance, identifying semiconductors with target band gaps may require band structure calculations for dozens or even hundreds of materials, while investigating the mechanical properties of two-dimensional materials necessitates the computation of elastic constants across various lattice types [1]. Such large-scale and systematic computational demands have driven the need for high-throughput first-principles calculations. By enabling batch job submission and systematic exploration of the materials parameter space, high-throughput approaches significantly accelerate the screening process and serve as a critical bridge between computational predictions and real-world applications [4].

Despite its broad recognition, the practical implementation of high-throughput computation still faces several challenges. First, batch job submission typically relies on complex scripting, which often lacks generality and must be tailored to specific types of calculations. Second, error handling and job monitoring during execution require substantial manual intervention. Third, the large volume of generated data necessitates specialized post-processing and analysis tools. Therefore, achieving truly automated and intelligent high-throughput computation remains an important research objective in the field of computational materials science.

In recent years, significant progress has been made in the development of high-throughput computation and automated workflow systems. Several software frameworks have been proposed for managing materials simulation workflows. AiiDA [5] is a highly extensible automation infrastructure that focuses on data storage, management, and automated workflow execution, offering full data provenance tracking and reproducibility. FireWorks[6] is a lightweight workflow management system designed specifically for scheduling and managing high-throughput computational tasks. QMflows [7] is a Python-based workflow framework that facilitates interoperability among different quantum chemistry software packages. DP-GENs[8] implement active learning loops for machine-learned interatomic potentials, accelerating molecular dynamics simulations. CatFlow[9] provides an automated workflow for training machine learning potentials, combining constrained molecular dynamics with active learning strategies to efficiently compute free energies of catalytic reactions. Mech2D[1], developed by a team at Hefei University of Technology, is a high-throughput computational tool for evaluating the mechanical properties of two-dimensional materials, capable of automatically calculating second-order elastic constants.

In the domain of AI-driven materials discovery, multi-agent systems have recently emerged as a promising research direction. TopoMAS [10] developed by the Chinese Academy of Sciences, is a multi-agent system for topological materials discovery. Through a hierarchical agent architecture, it coordinates multiple processes—including materials retrieval, knowledge reasoning, structure generation, and first-principles validation—thus enabling a complete workflow from user intent to materials discovery. Systems such as ChemCrow[11] and CACTUS[12] demonstrate the potential of large language models in orchestrating computational chemistry tools under scientific constraints. More recently, Ding et al. [13] introduced a decoupled agent–skill architecture for computational chemistry automation by using OpenClaw[14]. In this framework, general control logic is separated from domain-specific execution: planning skills translate scientific objectives into executable task specifications, while domain skills encapsulate concrete computational chemistry programs. DPDispatch handles task execution across heterogeneous high-performance computing environments. This architecture provides a new paradigm for constructing intelligent computational platforms.

Despite these advances, several limitations remain. First, most existing workflow systems rely on predefined task graphs and execution logic. Although some systems support limited runtime adaptability, their scalability becomes constrained when dealing with cross-software interoperability, context-dependent decision-making, or runtime fault recovery [14]. Second, while multi-agent systems in computational chemistry offer greater runtime flexibility, their workflow structures, recovery logic, tool routing strategies, and execution environment assumptions are often tightly coupled within specialized agent stacks. As a result, extending system capabilities typically requires redesigning the orchestration layer rather than simply replacing executable modules [14]. Furthermore, although current high-throughput tools excel at batch job submission, they remain limited in terms of intelligence. Traditional high-throughput computation essentially operates as a "data factory," focusing on large-scale task execution without the ability to dynamically adjust workflows based on intermediate results. In other words, existing systems are efficient at executing tasks but lack the capability to plan them intelligently—such as automatically generating task families based on scientific objectives or adaptively refining subsequent computations according to preliminary outcomes.

To address these challenges, this work proposes and implements an intelligent HTC (high-throughput computation) platform, HTC-Claw, built upon the OpenClaw framework. The key innovations of this platform are as follows:

**An agent-based framework for automatic task decomposition:** The system is capable of decomposing high-level research objectives into parallelizable and executable task sets. Upon receiving an exploratory query, the agent not only interprets its scientific intent but also autonomously generates task lists and optimizes submission strategies, enabling "one-command, full-family exploration."

**An intelligent workflow engine with closed-loop execution:** The platform integrates high-throughput computation with real-time analysis. The agent automatically triggers batch post-processing and analysis, proactively summarizes results, and presents them to users, transforming the workflow from a "submit–monitor" paradigm to a complete "submit–monitor–analyze–report" closed loop.

**Adaptive decision-making and workflow iteration:** The system incorporates conditional logic and iterative workflow capabilities, allowing it to make decisions based on intermediate results and autonomously initiate new computational tasks, forming a continuous "perception–decision–execution" cycle.

**Decoupled and modular intelligent scheduling system:** The platform features a modular design that decouples the scheduling system from functional modules, enabling flexible extension of capabilities. This architecture effectively mitigates errors in parameter configuration caused by agent hallucinations.

The remainder of this paper is organized as follows: Section 2 introduces the overall system architecture and workflow; Section 3 details the design and implementation of individual functional modules; Section 4 demonstrates the application of the system through representative case studies; and Section 5 concludes the paper and outlines future research directions.

## 2. System framework and workflow

### 2.1 Architecture design

The whole architecture is divided into user instruction layer, OpenClaw decision-making layer, and high-throughput computing platform layer, as shown in **Figure 1**. The user command layer is responsible for receiving natural language instructions from users and supports multiple input modalities, including direct conversational interaction, file uploads, and API calls. Users can describe their research objectives in natural language, such as "Evaluate the band gaps of all spinel structures," or "Search for corundum materials that retain metallic properties under a 2% strain"

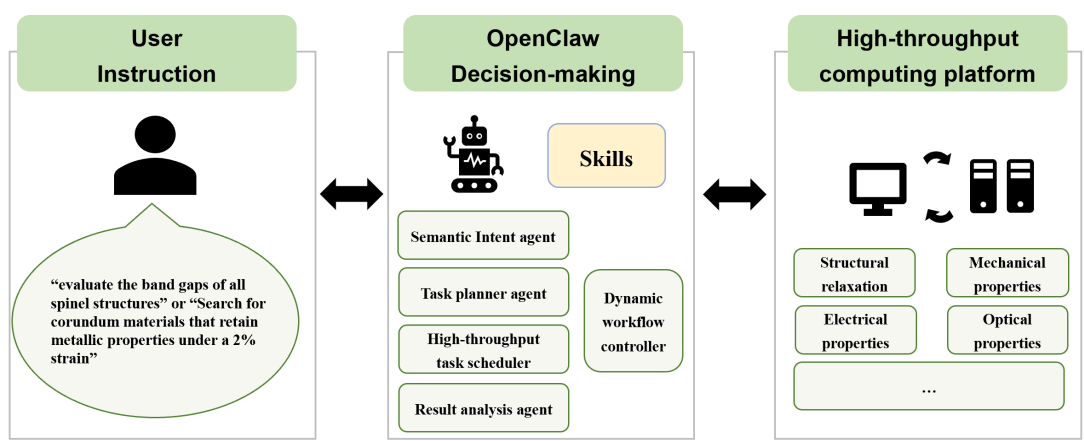

**Figure 1** HTC-Claw: OpenClaw-driven high-throughput materials discovery workflow.

The OpenClaw decision-making layer serves as the core intelligent hub of the system, integrating multiple specialized agents. The intent understanding agent parses user instructions and extracts key information, such as material types, target properties, and computational parameters. Based on the interpreted intent, the task planning agent automatically generates a complete list of tasks along with their execution sequence. The high-throughput task scheduler is responsible for the batch submission, monitoring, and management of computational jobs. Upon task completion, the result analysis agent performs automated data processing, result visualization, and knowledge extraction. Meanwhile, the dynamic workflow controller incorporates conditional logic and iterative capabilities, enabling it to adaptively adjust subsequent tasks based on intermediate results.

The high-throughput computing platform layer encapsulates a comprehensive set of computational functionalities, including structural optimization, electronic structure calculations, mechanical property evaluation, optical property analysis, thermal property prediction, molecular dynamics simulations, and machine-learning-assisted computations. The task dispatching system manages job submission and result retrieval across heterogeneous computing environments.

Furthermore, the intelligent design of the decision system and the modularization of the computational platform allow users to flexibly customize decision-making skills and extend computational functionalities. The decoupling of the scheduling system from the functional modules effectively mitigates the risk of incorrect parameter configurations arising from agent hallucinations.

**2.2 Core workflow**

The platform workflow follows a cyclic paradigm of "intent understanding → task planning → execution monitoring → result analysis." When a user submits a research objective, the system first employs the intent understanding agent to semantically parse the instruction, identifying the material types, target properties, and constraint conditions specified by the user. Subsequently, the task planning agent automatically generates a comprehensive task list based on the parsed information and assigns appropriate computational modules to each task.

For tasks requiring a high degree of parallelism, the high-throughput task scheduler packages and submits jobs in batches to computational clusters. During execution, the system continuously monitors task status and automatically handles common computational errors, such as convergence failures and structural instabilities. Upon completion of all tasks, the result analysis agent is triggered to perform batch processing of the outputs, generating aggregated reports and visualization plots.

For more complex tasks that require dynamic adaptation, the dynamic workflow controller intervenes in the execution process. It performs conditional evaluations based on the results of the initial round of computations, determining whether additional task iterations are necessary and how task parameters should be adjusted. This "perception–decision–execution" loop enables the system to handle complex exploratory workflows, rather than merely executing predefined static pipelines

**2.3 Agent Cooperation mechanism**

Multiple agents within the platform collaborate through message passing and shared state mechanisms. Each agent specializes in a specific category of tasks, while coordination among agents is achieved via a central orchestrator. When an agent requires assistance from other agents, it generates a task request and submits it to the coordinator, which subsequently dispatches the request to the appropriate agent for execution.

For example, when a user requests to "evaluate the band gaps of all spinel structures," the intent understanding agent first extracts the material type (spinel structures) and the target property (band gap). The task planning agent then queries the materials database to retrieve a list of candidate spinel structures and constructs a computational workflow for each material (i.e., structural optimization → static self-consistent calculation → band structure calculation). The high-throughput task scheduler subsequently submits these tasks in batch to the computational resources. Upon completion, the result analysis agent extracts the band gap data for all materials and generates comparative tables and trend visualizations. Throughout this process, the agents operate collaboratively, forming a fully automated workflow.

# 3 Function module and implementation

## 3.1 Intelligent high-throughput task planning

Intelligent high-throughput task planning constitutes one of the core innovations of this system. Unlike conventional batch submission approaches, intelligent task planning can automatically generate *task families* based on high-level research objectives specified by the user. The task planning agent is equipped with the following capabilities:

(1) **Goal decomposition**: When users propose exploratory objectives such as "evaluating the band gaps of all spinel structures," the agent can interpret the request and initiate queries to retrieve relevant crystal structures (e.g., spinel structures and inverse spinel structures) from materials databases.

(2) **Parameter space planning**: For tasks requiring systematic exploration of parameter spaces—such as varying alloy compositions or lattice constants—the agent can automatically generate comprehensive parameter combinations, thereby avoiding omissions.

(3) **Task dependency management**: The agent analyzes dependencies among tasks, automatically determines execution order, and constructs an optimal task topology.

Taking the example of "evaluating the band gaps of all spinel structures," the complete planning process of the agent is as follows:
- **Material retrieval**: Query the crystal structure database to obtain a list of candidate spinel materials.
- **Workflow construction**: For each material, generate a computational workflow: Module 101 (structural optimization) → Module 102 (static self-consistent calculation) → Module 401 (band structure calculation).
- **Task packaging**: Group tasks according to computational modules, forming multiple task batches.
- **Submission strategy optimization**: Determine optimal submission strategies based on available computational resources and task priorities.
- **Execution monitoring**: Monitor the status of each batch during execution and handle exceptions as they arise.

This intelligent planning capability eliminates the need for users to manually prepare individual computational tasks; instead, a single command can accomplish systematic calculations that would traditionally require days of manual effort.

In order to show this efficient way more intuitively, taking the above calculation of the bandgap of 3000 spinel structures as an example, we have made a simple and reasonable estimation of the time required for pure manual submission calculation and submission operation through the high-throughput computing platform, which is listed in **Table 1**

**Table 1** Comparison of Time Estimates for Manual vs. Automated Job Submission on a High-Throughput Computing Platform

| Metrics | Manual | HTC-Claw |
| --- | --- | --- |
| Task preparation time | ~2 day (1 min per structure) | ~1 min |
| Task submission time | ~5 hour (10 sec per structure) | ~10 sec |
| Data processing time | ~1 day (30 sec per structure) | ~30 sec |
| Total manual input | ~3 days | ~2 min |
| Queue waiting time | Take time | Efficiency |

### 3.2 Dynamic Workflows and Adaptive Exploration

The dynamic workflow controller is a key component for enabling "intelligent high-throughput" computation. It provides the following core functionalities:

(1) **Conditional triggering mechanisms**: The system can trigger subsequent tasks based on intermediate computational results. For example, after completing the first round of elastic constant calculations, materials satisfying mechanical stability criteria can be automatically selected for subsequent electronic structure calculations.

(2) **Workflow iteration capability**: When preliminary results do not meet expectations, the system can automatically adjust task directions and perform targeted exploration. This "learning-while-computing" paradigm significantly improves search efficiency.

(3) **Multi-stage task coordination**: Complex problems may require multiple cycles of computation–analysis–decision. The dynamic workflow controller can manage such long-range dependencies.

A representative example of adaptive exploration is the task of "identifying spinel materials that remain metallic under 2% strain":

- **First round**: The agent selects candidate materials from the database and submits parallel calculations of elastic constants (Module 601) and initial electronic structures (Module 102).
- **First-round analysis**: The agent automatically extracts elastic constants, applies a 2% strain screening criterion, and identifies candidates that satisfy mechanical requirements.

- **Second round**: For the filtered materials, the system automatically submits electronic structure calculations under applied strain to verify whether metallicity is preserved.
- **Result aggregation**: A final list of materials satisfying both mechanical stability and metallicity criteria is generated.

This condition-triggered dynamic task expansion mechanism addresses the limitation of traditional high-throughput approaches, where data generation is efficient but knowledge extraction is slow. In this framework, the agent functions not only as a task initiator but also as an initial data analyst.

**3.3 Batch Task Scheduling and Error Recovery**

The high-throughput task scheduler is responsible for the batch submission and management of large-scale computational tasks. It is tightly integrated with the high-throughput computing platform and supports task execution across heterogeneous computational environments.

The core functionalities of the scheduler include:

(1) **Task packaging and batch submission**: Many similar tasks are grouped into task sets and submitted in batches to computational clusters, thereby improving scheduling efficiency.

(2) **Load balancing**: Tasks are dynamically distributed based on the current workload of computing nodes and task characteristics, optimizing resource utilization.

(3) **Error monitoring and automatic recovery**: The system continuously monitors task execution status. For common computational errors—such as self-consistent field (SCF) convergence failures or unreasonable ionic configurations—it performs automatic error diagnosis and attempts recovery. For errors that cannot be resolved automatically, the system records detailed diagnostic information and notifies the user.

(4) **Automatic result retrieval**: Upon task completion, output files are automatically collected and subjected to preliminary integrity checks to ensure data usability.

This condition-triggered dynamic task expansion mechanism addresses the fundamental limitation of traditional high-throughput computation, where data generation is efficient but knowledge extraction is comparatively slow. Within this framework, the agent functions not only as an initiator of computational tasks but also as a first-level data analyst responsible for preliminary screening.

(2) **Load balancing**: Tasks are intelligently distributed based on the current workload of computing nodes and task characteristics, optimizing resource utilization.

(3) **Error monitoring and automatic recovery**: The system continuously monitors task execution status. For common computational errors (e.g., self-consistent field (SCF) convergence failures or unreasonable ionic configurations), it performs automatic error diagnosis and attempts corrective actions. For errors that cannot be automatically resolved, the system records detailed error information and notifies the user.

(4) **Automatic result retrieval**: Upon task completion, output files are automatically collected and subjected to preliminary integrity checks to ensure data usability.

## 3.4 Result Analysis and Knowledge Extraction

The result analysis agent is automatically activated upon completion of computations, performing large-scale data processing and knowledge extraction from extensive simulation outputs. Its primary functionalities include:

(1) **Batch data extraction**: Automatically extracts key physical quantities—such as band gaps, elastic constants, and formation energies—from large volumes of computational output files.

(2) **Statistical analysis**: Performs statistical analyses on the extracted data, including the calculation of mean values, variances, and extrema, as well as the identification of outliers.

(3) **Visualization generation**: Automatically produces visualization outputs such as comparative tables, trend plots, and distribution diagrams.

(4) **Knowledge association**: Links computational results with materials databases and existing literature to uncover potential structure–property relationships.

For example, for a set of band structure calculations, the result analysis agent automatically generates a comparative table of band gaps, as show in **Figure 2**, ranks materials according to band gap values, identifies candidates with potential applications (e.g., narrow-gap or wide-gap materials), and produces trend plots correlating band gap with material composition.

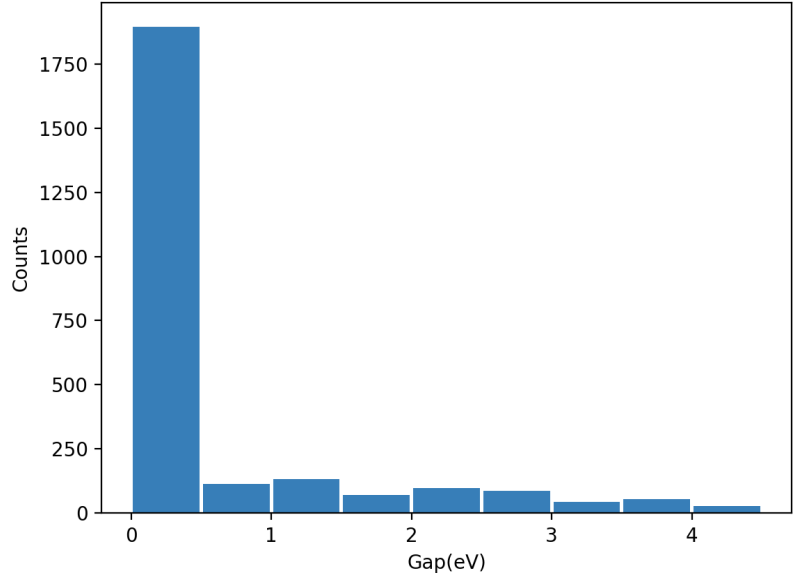

**Figure2.** The calculated bandgap distribution of spinel structures

### 3.5 Functional modules of high-throughput computing platforms

The platform encapsulates complete first principles computing functional modules, as shown in **Table 2,** covering the main research scenarios of material computing and supporting customized module extensions:

**Table 2.** First-Principles Computational Modules Supported by the Platform

| Module number | Features | Related Software |
|---|---|---|
| 101/102 | Structural relaxation and static self-consistent calculations | VASP |
| 201/202 | Batch job submission and management | Slurm |
| 301-303 | Optical properties calculation | VASP MTP-FIRE |
| 401-404 | Electrical properties calculation | VASP pymatgen |
| 501-503 | Thermal properties calculation | VASP Phonopy MTP |
| 601-602/701-706 | Mechanical properties calculation | VASP Lammps |
| 801-802 | Finite element simulation | abaqus |
| 901-905 | Machine learning assisted reverse design of crystal structures | GNN CDVAE |
| … | … | … |

These modules interface with the OpenClaw agent layer through a unified interface, allowing agents to flexibly invoke required modules according to task requirements and assemble them into complete workflows.

# 4. Case Study

**Adaptive Exploration for Metallic Behavior under Strain**

This case study illustrates the system's **dynamic workflow** capabilities, as shown in **Figure 3.** The user instruction is: *"Identify spinel materials that remain metallic under 2% strain."*

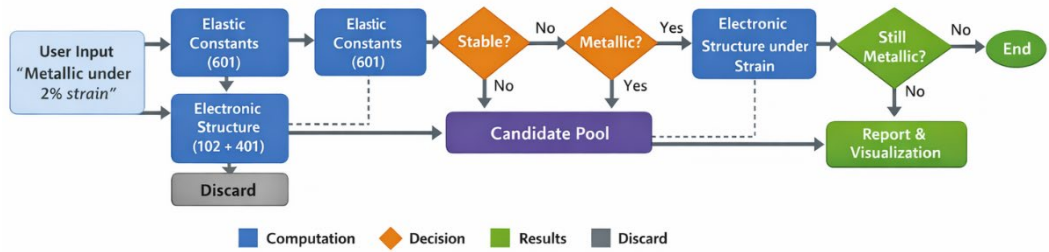

**Figure 3.** Workflow of the adaptive exploration for metallic behavior under strain

**Agent execution process:**

**First round: initial screening of candidate materials**

- Retrieve candidate spinel structures from the materials database.
- Submit batch structural relaxation calculations (Module 101).
- Submit batch calculations of elastic constants (Module 601) and electronic structures (Modules 102 + 401).
- Await completion of all computations.

**First-round analysis: mechanical stability screening**

- Extract elastic constants for all materials.
- Apply Born stability criteria to filter mechanically stable materials:

$$C_{11} > 0, C_{66} > 0 \quad (1)$$
$$C_{11}C_{22} > C_{12}^2 \quad (2)$$

**Second round: metallicity under strain evaluation**

- Apply 2% strain to the filtered materials.
- Submit electronic structure calculations under strain (Modules 102 + 401).
- Examine band gaps to determine whether metallicity is preserved.

- Aggregate results and generate a report listing materials that satisfy all criteria, including band gap–strain relationship plots and detailed calculation reports for each candidate.

This case demonstrates the agent's ability to **make decisions based on intermediate results** and autonomously initiate new computational tasks. The workflow incorporates decision branches and iterative loops, forming an intelligent perception–decision–execution closed loop that far surpasses traditional static batch submission processes.

## 5. Conclusion

In this work, we propose and implement HTC-Claw, an intelligent high-throughput *ab initio* computational platform based on OpenClaw. By deeply integrating OpenClaw's agent-based architecture with a high-throughput computing platform, HTC-Claw realizes a fully automated, interpretable, and adaptive workflow that spans from user natural language instructions to cross-scale simulations and data generation. The main contributions of this work are as follows:

**Agent framework for goal decomposition and parallel task planning:** When users specify high-level research objectives, the agents automatically interpret their scientific intent, generate task families, and optimize submission strategies, achieving *"one command, full-family exploration."*

**Intelligent workflow engine enabling a high-throughput computation–analysis closed loop:** Upon task completion, agents automatically trigger batch post-processing and analysis, upgrading the workflow from *"submit–monitor"* to a complete *"submit–monitor–analyze–report"* loop.

**Condition-triggered dynamic task expansion mechanism:** Agents can make decisions based on intermediate results and autonomously initiate new tasks, implementing a perception–decision–execution cycle.

**Intelligent decision system and modular computational platform:** The platform allows users to freely customize decision-making skills and extend computational modules. Decoupling of the scheduling system from functional modules effectively mitigates errors in parameter configuration caused by agent hallucinations.

Validation through representative case studies demonstrates that HTC-Claw significantly enhances computational efficiency, reduces manual intervention, and enables intelligent exploration in materials research. Future work will focus on further integrating knowledge graphs, expanding computational modules, and exploring more advanced forms of multi-agent collaboration.

## Acknowledgments

The research was supported by the National Natural Science Foundation of China (Grants U23A20565 and 52301194), Innovation Program of Shanghai Municipal Education Commission (Grant 2023ZKZD15), the Shanghai Science and Technology

Commission (Grant 22511100400), the startup funding from Shanghai Jiao Tong University (WH220405009), and open research fund of Suzhou Laboratory (Grants SZLAB-1108-2024-TS003). The computations in this paper were run on the Siyuan-1 cluster supported by the Center for High Performance Computing at Shanghai Jiao Tong University.

## Supporting information

This project can be found in https://github.com/ldzeng/HTC-Claw